Title: Leveraging latent persistency in United States patent and trademark applications to gain insight into the evolution of an innovation-driven economy

Author: Iraj Daizadeh, PhD, Global Regulatory Affairs, Takeda Pharmaceuticals, 40 Landsdowne Street, Cambridge, MA 02139.  All correspondence should be sent to: iraj.daizadeh@takeda.com

Abstract:

Objective: An understanding of when one or more external factors may influence the evolution of innovation tracking indices (such as US patent and trademark applications (PTA)) is an important aspect of examining economic progress/regress. Using exploratory statistics, the analysis uses a novel tool to leverage the long-range dependency (LRD) intrinsic to PTA to resolve when such factor(s) may have caused significant disruptions in the evolution of the indices, and thus give insight into substantive economic growth dynamics.

Approach: This paper explores the use of the Chronological Hurst Exponent (CHE) to explore the LRD using overlapping time windows to quantify long-memory dynamics in the monthly PTA time-series spanning 1977 to 2016.

Results/Discussion: The CHE is found to increase in a clear S-curve pattern, achieving persistence (H~1) from non-persistence (H~0.5). For patents, the inflection occurred over a span of 10 years (1980-1990), while it was much sharper (3 years) for trademarks (1977-1980).

Conclusions/Originality/Value: This analysis suggests (in part) that the rapid augmentation in R&D expenditure and the introduction of the various patent directed policy acts (e.g., Bayh-Dole, Stevenson-Wydler) are the key impetuses behind persistency, latent in PTA. The post-1990's exogenic factors seem to be simply maintaining the high degree and consistency of the persistency metric. These findings suggest investigators should consider latent persistency when using these data and the CHE may be an important tool to investigate the impact of substantive exogenous variables on growth dynamics.

Introduction: Literature Review and Problem Statement

Intellectual property (IP)-based metrics (initially patents and more recently trademarks) have been used as proxy measures of global politico-socio-economic innovative behavior [on the micro (firm)-, meso (research institution, cluster)-, and/or macro (country/regional)-level] for decades (see, e.g., Dziallas and Blind, 2019). These and related indices focus on the time-dependent ebbs and flows of absolute counts of observations and/or derivates thereof. Importantly, it is understood that these data integrate one or more extrinsic forces and/or factors that are either directly or indirectly influenced the chronological evolution of these data in some way (see, e.g., Daizadeh, 2007, 2009, 2021). Indeed, a vast majority of this research either test or generate hypothesis of how such extrinsic impulses – such as promulgation of a specific policy framework, gross effects of research and development expenditures, sector specific technology dynamics (e.g., disruptive versus incremental) – may affect the time course of these IP-metrics (see, e.g., Grimaldi and Cricelli, 2020; Flikkema, et al., 2015; Daizadeh, 2007, 2021). Unfortunately, to the author's knowledge, there is limited-to-no inquiries exploring the intrinsic make-up of these time series data.

Even in the simplest of cases, the interpretative potential of time series data, such as those of patents and trademarks, may be challenging, however, due to their intrinsic behavior (s) (see, e.g., Cheng et al., 2015). Importantly and simultaneously, some of these properties are also of interest due to the potential to elucidate interesting dynamics. For example, memory characteristics across length scales (short, medium, and long) suggest the existence of various economic cycles (see, e.g., Alvarez-Ramirez et al., 2020). Concordantly, these same memory characteristics can obfuscate true signal detection, due to – for example – mistakenly considering long-memory and non-stationary complex dynamics (see, e.g., Saha et al., 2020). Other properties such as linearity, normality, and seasonality of the time series may further exacerbate mis-interpretations. Statistical time series practitioners exert great effort in developing methods that would increase the certitude of the interpretation by accommodating one or more of these properties. Resolving the suite of issues is the first step in establishing robust statistical approaches and increased confidence in model interpretability.

The Hurst exponent is statistical time-series tool that has been used to better understand memory effects in time series data and has been applied to various fields including: earth sciences (see, e.g., Slino, et al., 2020), economics (see, e.g., Wu and Chen, 2020), and others. While there are other ways, the Hurst exponent (a measurement of memory) is classically defined as $H \sim \ln(R / S)_t / \ln(t)$, where R and S is the rescaled range and standard deviation, respectively, and t is a time window. An H=0.5, an H<0.5, and an H>0.5 indicates a random walk (non-persistent), an anti-persistent, and a persistent (trend reinforcing) time series behavior, respectively (see, e.g., Mandelbrot and Wallis, 1968, 1969).

To the author's knowledge, there have not been any investigations of the Hurst exponent with regards to either patents or trademarks. In terms of patents, however, some elements of memory effects have been explored in the context of economic cycles by several authors (see, e.g., Korotayev, 2011; Haustein and Neuwirth, 1982, Alvarez-Ramirez, 2020, Epicoco, 2020, Daizadeh, 2021a). Korotayev et al (2011) demonstrated a Kondratieff (K)-wave pattern when investigating the dynamics of the annual number of global patents per million population from 1900 to 2008. As summarized by (Alvarez-Ramirez, 2020), Korotayev showed that patents presented with "a steady increase during the upswing phase of Kondratieff's cycle, and a pronounced decrease during the downswing phase." Haustein and Neuwirth (1982) found that "industrial production on patents with a lag of 9 years." More recently, Epicoco (2020) fitted the information and communications technologies cycle with that of the economic using patent and productivity data and proposed "the current productivity slowdown may be a signal that the economic system needs to change its leading technologies."

Carbone, Castelli, and Stanley (2004) proposed a time dependent Hurst exponent based on a detrending moving average (DMA). Here, the Hurst exponent is a log-log slope of the DMA standard deviation against window (see equations 1 and 2 therein). The authors conclude from an analysis of artificial and observed time series of financial data that the time-variability is much "richer" than anticipated from a mono-fractal approach. To accommodate non-stationary effects, the work was subsequently extended with the aid of detrended fluctuation analysis over non-overlapping window lengths by Alverez-Ramirez and colleagues (2020). Effectively, these approaches to time-dependent Hurst exponent calculations are model parameterized.

In this work, and as described in the Methodology section below, a non-model parameterized chronological Hurst exponent (CHE) is proposed that when applied to a given time series may identify significant changes in the persistency of memory. The method is straightforward to implement since it simply uses a standard estimate of the Hurst exponent calculated from the initial time point to month $\alpha$, where $\alpha$ is a monthly increment. The output of the CHE calculation is described as the time series plot of each of the Hurst exponents, allowing a qualitative view of the Hurst exponent over a given time period (see Methods). This approach allows for an arbitrary method to calculate the Hurst exponent, while taking into regards the nuances of the time series. Here, the method is applied to US patent and trademark applications (PTA) from 1977-2016. The date range selected is chronologically broad (a period of over 40 years) and used in prior work, and thus presented here. Interested readers may extend the data range accordingly, as all data and R Programs used for this manuscript are available (Appendix; Daizadeh, 2021b).

As described in the Results section below, from the application of this novel tool to PTA, it is found that the CHE evolves in a highly descriptive and idiosyncratic S-like pattern: from non-persistent ($H\sim0.5$) to saturation (trend reinforcing persistent ($H\sim1$) level via a quickly evolving inflection period (see Figures 1-3). For patents, the inflection occurred over a span of 10 years (1980-1990), while it was much sharper (3 years) for trademarks (1977-1980). As will be further discussed below, these findings suggest: investigators should consider latent persistency when using these data; exogenous factors after the identified inflection points for these indices have only incrementally strengthened intrinsic memory; and, the CHE may be an important tool to investigate the impact of substantive exogenous variables on growth dynamics. Qualitative correlation of the timing of the inflection points for patents and trademarks suggest the importance of research and development expenditure.

Methodology:

The data were comprised of the monthly number of US patent applications (Patents) and the monthly number of US trademarks filings (Trademarks) (together, PTA) from 1977 to 2016, and were obtained from the United States Patent and Trademark Office (USPTO) as described below:

- Patents:
    - Website: http://patft.uspto.gov/netahtml/PTO/search-adv.htm
    - Search pattern: Application Filing Date: "APD/MM/$/YYYY"
- Trademarks:
    - Website: http://tmsearch.uspto.gov/bin/gate.exe?f=tess&state=4804:57thz4.1.1
    - Search pattern: Filing Date: "(YYYYMM$)[FD]"

Note: MM/YYYY is the 2/4 digital representation for month/year.

The two searches resulted in 472 datapoints – representing monthly observations over the period of study (approximately 40 years) – for each variable and imported into R for processing (Appendix; Daizadeh, 2021b).

Methodology followed standard implementation, and default parameters were used throughout. The general algorithm for the analysis is as follows:

- Load time series (R package 'tseries' (Trapletti and Hornik, 2019)), identify and replace outliers with average of prior and posterior-month values (R package 'tsoutliers' (López-de-Lacalle, 2019). Note: 3 outliers were determined for Trademarks (September 1982; November 1989; and June 1999) and 4 for Patents (September 1982, June 1995, October 2007, and March 2013).

- Calculate descriptive statistics [including standard deviation, kurtosis, and skew (R package 'moments' (Komsta and Novomestky, 2015)] and auto/serial correlation (base R package).

- Calculate intrinsic variables: normality (R package 'nortest' (Gross and Ligges, 2015)), stationarity (R package 'forecast' (Hyndman, et al., 2020; Hyndman and Khandakar, 2008; R package 'aTSA' (Qiu, 2015)), seasonality (R package 'seastests' (Ollech, 2019)), and non-linearity (R package 'nonlinearTseries' (Garcia, 2020))

- Calculate chronological Hurst exponent: Determine Hurst exponent based on Hyndman implementation (R package 'tsfeatures' (Hyndman et al, 2020)) using the following algorithm:

  o for (i in start:end) { hurst-IP[i] <- hurst (time[1:(1+i*1)]) }, where IP is either Trademarks or Patents; start = September, 1977; end = December 2016; I = monthly increments

  o Note: The Hyndman approach – one of several methods to calculate (estimate) the Hurst Exponent (Shang, 2020) – is defined as 0.5 plus the maximum likelihood estimation of the fractional differencing order (see Hyndman et al, 2020); thus, it has properties that differ than Hurst's original definition (e.g., no singularities at certain scales). In principal, any approach should produce qualitatively the same result as that outline above, albeit additional work is required to confirm the approaches sensitively.

Results and Discussion:

*General Statistics*

Generally, the time series were similar in structure with a general cobra-like structure (see upper graphs of Figures 1, 2, and 3) and similarities in shape of the distributions (e.g., approximately symmetric (skew) and platykurtic) (see Table 1). The time series showed clear long-memory tendency as presented in the auto and serial correlation functions with lag much great than 2. Lastly, both time series were non-normal, non-stationary [with a single difference ((t-1) – t) bringing them into stationarity – that is, integration of order 1 typical of econometric data], seasonal, and non-linear (see Table 2).

Figure 1: The monthly number of Patents (top graph) from 1977-2016 with its corresponding chronological Hurst values (bottom graph)

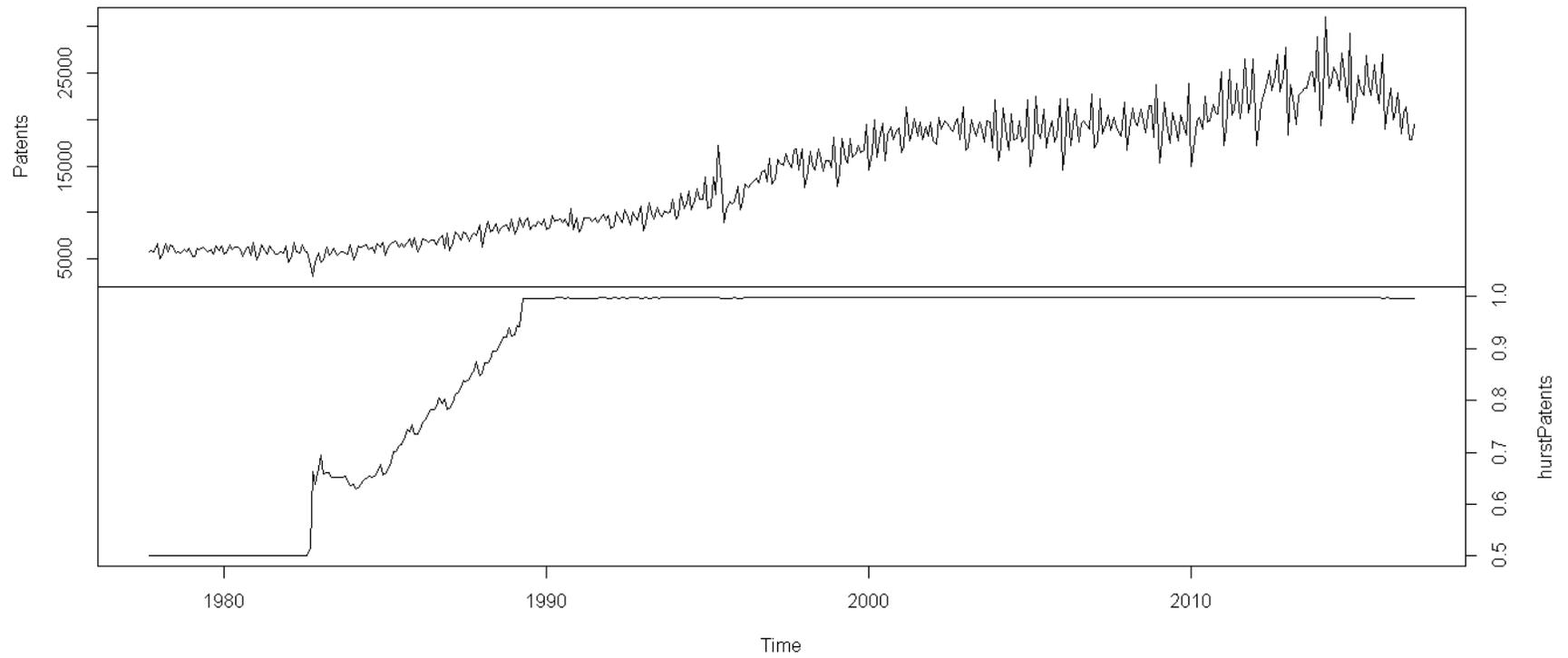

Figure 2: The monthly number of Trademarks (top graph) from 1977-2016 with its corresponding chronological Hurst values (bottom graph)

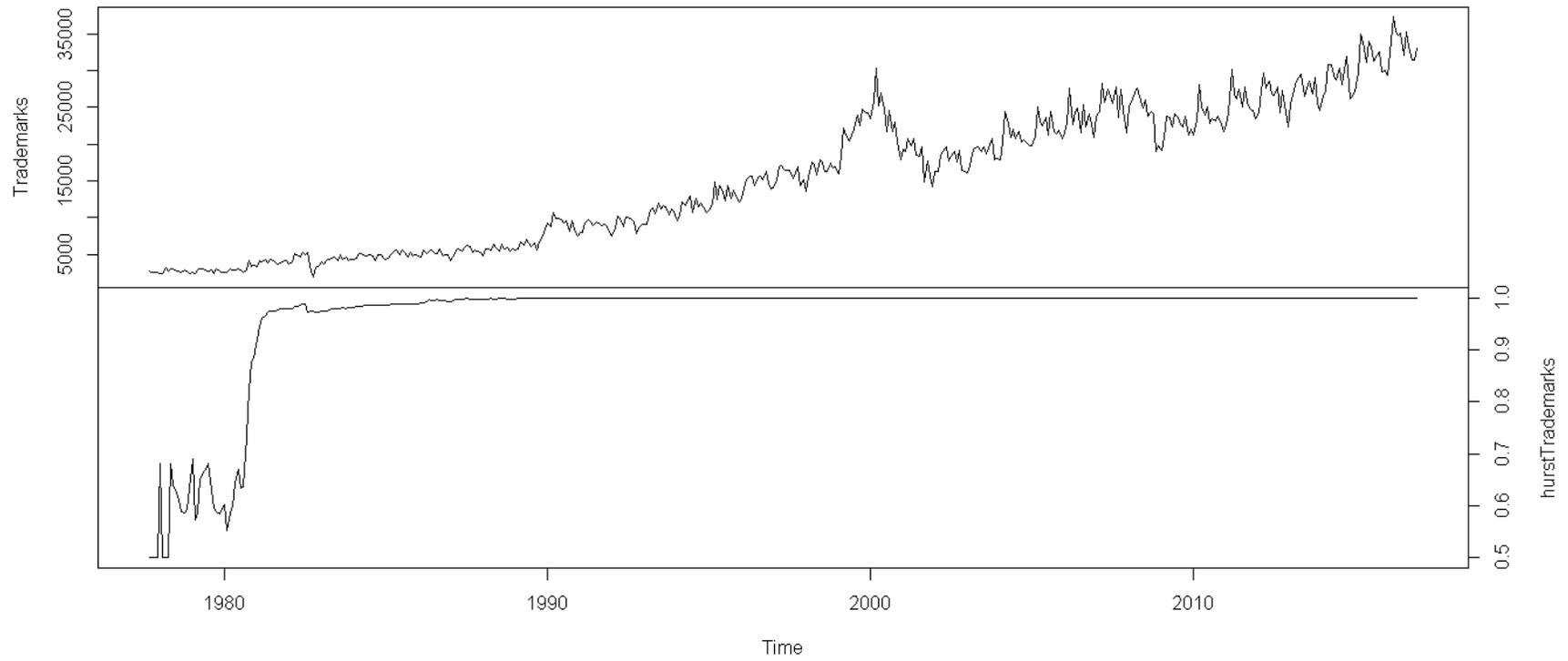

Figure 3: Comparison of chronological Hurst values between Patents and Trademarks

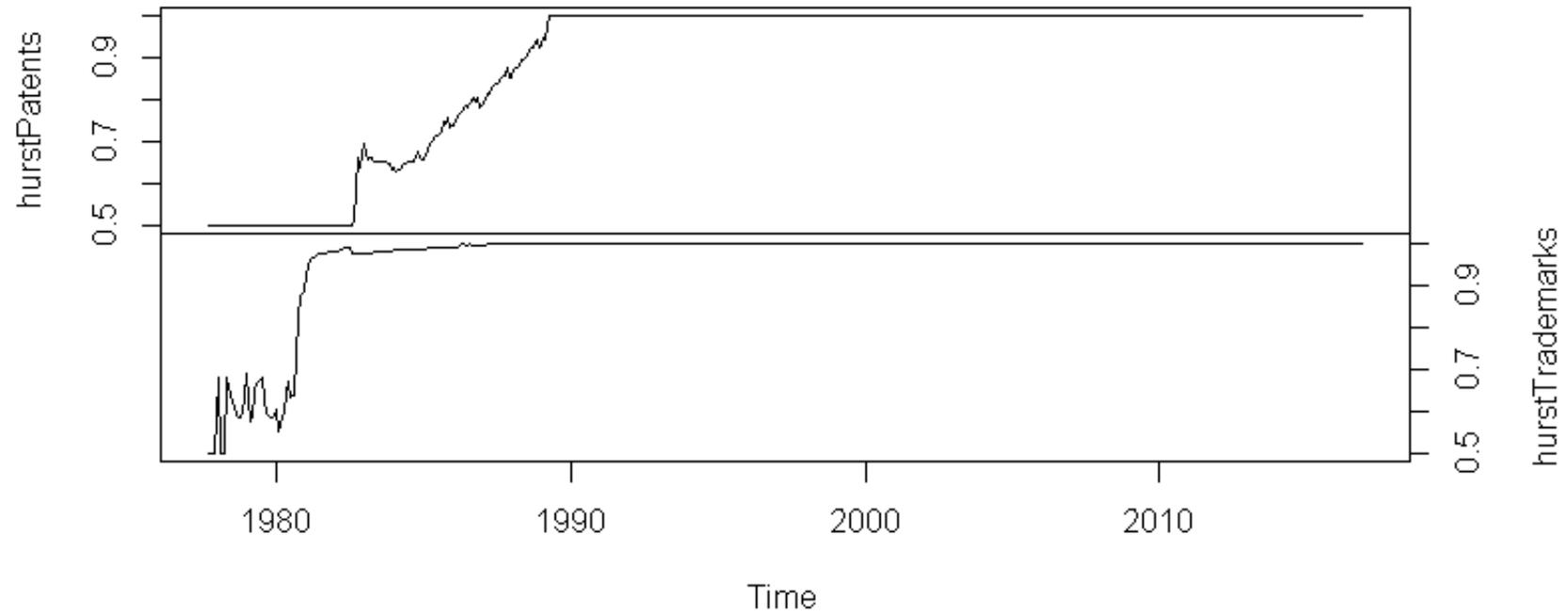

Table 1: Descriptive statistics of Patents and Trademarks

|  | Patents | Trademarks |
|---|---|---|
| Minimum | 3134 | 1895 |
| 1$^{st}$ Quantile | 7598 | 5537 |
| Median | 14468 | 15456 |
| Mean | 13930 | 15276 |
| 3$^{rd}$ Quantile | 19422 | 23574 |
| Maximum | 30969 | 37317 |
| Stand Deviation | 6523.347 | 9418.054 |
| Skewness | 0.1854975 | 0.2020644 |
| Kurtosis | 1.757654 | 1.763723 |

Table 2: General results of intrinsic parameters of Patents and Trademarks (see Appendix; Daizadeh, 2021b)

|  | Patents | Trademarks |
|---|---|---|
| Normality | Non-normal | Non-normal |
| Stationarity | Non-stationary (order of integration: 1) | Non-stationary (order of integration: 1) |
| Seasonality | Seasonal | Seasonal |
| Non-linearity | Non-linear | Non-linear |

*Chronological Hurst Exponent*

Examining Figures 1 and 2, the time evolution of PTA shows a general trend-reinforcing pattern, edging generally up with time. There are variations of course partly due to the identified seasonal effects. The Chronological Hurst Exponent (CHE) correspondingly evolves, with non-persistence (H~0.5) near the beginning of the time series (notice: this is where the absolute counts of the PTA are effectively static). As the time series evolves, the CHE quickly jumps to saturation (H~1). A Hurst exponent at such a high relative value suggests that that the trend will continue to persist in the original time series – that is, the exponent reflects the similarity of prior absolute values of PTA to future values; in other words: as prior timepoints values edge up, it is likely that the next value will also be up. This is consistent with the long auto and serial correlations in the original time series.

Co-positioning PTA CHE (as in Figure 3), a similarity in S-structure is presented.  The scaling regimes of are trichotomized in both time series into the following periods: (1) non-persistence (albeit for Trademarks mild persistence is observed with H values oscillating between 0.5 and 0.7); (2) time-varying persistence (an inflection, where H slopes from 0.5/0.6 to near 1); and, (3) persistence (H~1). Specifically, it is found that:

- Period 1: Weak to no persistence prior to 1980s:
    - Patents: The flat H exponent, as shown in Figure 1 (lower graph) and Figure 3 (upper graph), reflects the neutral acceleration in the time series.
    - Trademarks: The variation in H exponent (~0.5 to ~0.7) reflects a positive trajectory as depicted in Figures 2 (upper graph) and 3 in the reference time series.
- Period 2: A time-varying Hurst exponent up to and during the 1980s and 1990s

- o Patents: The value of the H exponent began to grow with a rapid linear rate of approximately 0.1-0.2 Hurst values per year, commencing with an explosive charge in the mid-1980s, and terminating with saturation before 1990.
- o Trademarks: The value of the H exponent grew at an exceptional rate at roughly 0.3 over two years, from effective baseline to near saturation by 1983.
- Period 3: Saturated persistence for trademarks beginning by 1983 and for patents in the 1990s throughout the duration of the reporting period 2016.

Focusing on the time-varying Period 2, while more work needs to be done to better our understanding of the dynamics in the 1980s and 1990s that affected these proxy measures of innovativeness, some thoughts present themselves for potentially testable causality hypotheses:

- Economic: Total spending on research and development grew from $60B USD in 1975 to $100B by 1985, concomitantly while the contribution from industry to the percent of Gross Domestic Product rose from 1% to nearly 1.5% during that same time (see Figure 1 in Hunt, 1999). Indeed, patent activity (number of Patent Applications) also doubled grew from 100k (1980) to 200k (2000) (see Figure 2 in *ibid*). Could industrial R&D expenditures directly drive persistency of Patents and Trademarks?
- Policy: There were significant competitive policy initiatives during the 1980s including: the Bayh-Dole Act (Patent and Trademark Law Amendments Act (Pub. L. 96-517, December 12, 1980), Public Law 96-480, Stevenson-Wydler Technology Innovation Act (as amended in 1986 and 1990), and others (see Table 1 in Coriat and Orsi, 2002). Was the persistency initiated by one or more of these policies (including, for example, new patentability types: software and/or business plans)?

- Technology dynamics: S-shaped growth for individual technological innovations may contribute to the overall Patent evolution dynamic (see, e.g., Anderson, 1999). A linear combination of such patent growths may be affecting the persistence of the overall Patents structure. To the author's knowledge, there is limited to no information on any S-shaped distribution for Trademarks, a potential for future query. Do technology dynamics associated with innovation cumulatively affect persistence in the innovation metrics?

Conclusion:

The goal for the use of a novel method – termed the chronological Hurst exponent (CHE) – to investigate long-term memory dynamics in time series data is to describe the overall persistency tend and, importantly, to identify the key dates in which there is a change in persistency (anti-persistence, non-persistence, persistence). Here, the CHE was applied to two innovative-tracking, intellectual property-driven data: the monthly numbers of patent and trademark applications. CHE is a simple to use method, trivially constructed in any programming language (Appendix; Daizadeh, 2021b) and easily applied to longitudinal data. Additional work is required to better understand the approach, including using different memory quantification tools (e.g., different versions of the Hurst exponent calculation) and different (longer) datasets with different characteristics.

In summary, this paper presents the first use of the CHE as well as initial results of its application to PTA. The importance of persistency as an intrinsic factor of time series data should not be underestimated as effects such as trend-reinforcing behavior may lead to biased results when interpreting time series data, as momentum of the time series may cover salient investigative concerns (recall that the Hurst exponent is a measure of self-similarity). It is a hope of this work that investigators continue to examine the innovative accomplishments of the late 20th century as well as consider persistency measures in current assessments of techno-economic progress.

Conflict of Interest Statement: :

The author is an employee of Takeda Pharmaceuticals; however, this work was completed independently of his employment. The views expressed in this article may not represent those of Takeda Pharmaceuticals.

Statement of Data Consent:

The data generated during the development of this study has been included in the manuscript.

References:

Alvarez-Ramirez, J.; Rodriguez, E.; Ibarra-Valdez, C. (2020) Medium-term cycles in the dynamics of the Dow Jones Index for the period 1985–2019. Physica A: Statistical Mechanics and its Applications. 546: 124017 https://doi.org/10.1016/j.physa.2019.124017

Anderson, B. (1999) The hunt for S-shaped growth paths in technological innovation: a patent study. J Evol Econ (1999) 9: 487-526 https://doi.org/10.1007/s001910050093

Carbone, A.; Castelli, G.; Stanley, H.E. (2004) Time-dependent Hurst exponent in financial time series. Physica A: Statistical Mechanics and its Applications. 344(1-2): 267-271

https://doi.org/10.1016/j.physa.2004.06.130

Cheng, C., Sa-Ngasoongsong, A., Beyca, O., Le, T., Yang, H., Kong, Z., & Bukkapatnam, S. T. (2015). Time series forecasting for nonlinear and non-stationary processes: A review and comparative study. Iie Transactions, 47(10), 1053-1071.

Coriat, B.; Orsi, F. (2002) Establishing a new intellectual property rights regime in the United States: Origins, content and problems. Research Policy 31:1491-1507. https://doi.org/10.1016/S0048-7333(02)00078-1

Daizadeh, I. (2007) Issued US patents, patent-related global academic and media publications, and the US market indices are inter-correlated, with varying growth patterns. Scientometrics 73(1): 29-36.

https://doi.org/10.1007/s11192-007-1749-1

Daizadeh, I. (2009) An intellectual property-based corporate strategy: An R&D spend, patent, trademark, media communication, and market price innovation agenda. Scientometrics 80(3): 731-746.

Daizadeh, I. (2021). Trademark and patent applications are structurally near-identical and cointegrated: Implications for studies in innovation. *Iberoamerican Journal of Science Measurement and Communication*, 1(2). https://doi.org/10.47909/ijsmc.33

Daizadeh, I. (2021a). US FDA Drug Approvals are Persistent and Polycyclic: Insights into Economic Cycles, Innovation Dynamics, and National Policy. *Therapeutic Innovation and Regulatory Science*.

https://doi.org/10.1007/s43441-021-00279-8

Daizadeh, I. (2021b). Chronological Hurst exponent elucidates latent persistency within patents and trademarks applications reflecting strength of innovation initiatives between 1977 and 2016. arXiv preprint arXiv:2101.02588.

Dziallas, M.; Blind, K. (2019) Innovation indicators throughout the innovation process: An extensive literature analysis.  Technovation 80-81: 3-29.

Epicoco, M. (2020) Technological Revolutions and Economic Development: Endogenous and Exogenous Fluctuations. J. Knowl. Econ.. https://doi.org/10.1007/s13132-020-00671-z

Garcia, C.A. (2020) nonlinearTseries: Nonlinear Time Series Analysis. R package version 0.2.10.

https://CRAN.R-project.org/package=nonlinearTseries

Grimaldi, M.; Cricelli, L (2020) Indexes of patent value: a systematic literature review and classification, Knowledge Management Research & Practice, 18:2, 214-233, DOI:10.1080/14778238.2019.1638737

Haustein H.D.; Neuwirth E. (1982) Long waves in world industrial production, energy consumption, innovations, inventions, and patents and their identification by spectral analysis. Technol. Forecast. Soc. Change, 22:53-89 https://doi.org/10.1016/0040-1625(82)90028-2

Hunt, R.M. (1999) Patent reform: a mixed blessing for the US economy. Business Review. Federal Bank of Philadelphia, November-December. Accessed via https://www.philadelphiafed.org/-/media/research-and-data/publications/business-review/1999/november-december/brnd99rh.pdf

Hyndman, R.; Kang, Y.; Montero-Manso, P.; Talagala, T.; Wang, E.; Yang, Y.; O'Hara-Wild, M. (2020) tsfeatures: Time Series Feature Extraction. R package version 1.0.2. https://CRAN.R-project.org/package=tsfeatures

Komsta, L.; Novomestky, F. (2015). moments: Moments, cumulants, skewness, kurtosis and related tests. R package version 0.14. https://CRAN.R-project.org/package=moments

Gross, J.; Ligges, U. (2015). nortest: Tests for Normality. R package version 1.0-4. https://CRAN.R-project.org/package=nortest

Hyndman, R.; Athanasopoulos, G.; Bergmeir, C.; Caceres, G.; Chhay, L.; O'Hara-Wild, M.; Petropoulos, F.; Razbash, S.; Wang, E.; Yasmeen, F. (2020). forecast: Forecasting functions for time series and linear models. R package version 8.12, http://pkg.robjhyndman.com/forecast

Hyndman, R.J.; Khandakar, Y. (2008). "Automatic time series forecasting: the forecast package for R." *Journal of Statistical Software*, *26*(3), 1-22. http://www.jstatsoft.org/article/view/v027i03

Korotayev, A.; Zinkina, J.; Bogevolnov, J.; (2011) Kondratieff waves in global invention activity (1900–2008). Technol. Forecast. Soc. Change, 78:1280-1284 https://doi.org/10.1016/j.techfore.2011.02.011

López-de-Lacalle, J. (2019). tsoutliers: Detection of Outliers in Time Series. R package version 0.6-8. https://CRAN.R-project.org/package=tsoutliers

Mandelbrot, B.B.; Wallis, J.R. (1969) Noah, Joseph, and Operational Hydrology. Water Resources Research 4(5) https://doi.org/10.1029/WR004i005p00909

Mandelbrot, B.B.; Wallis, J.R. (1969) Robustness of the rescaled range R/S in the measurement of noncyclic long run statistical dependence. Water Resources Research 5(5)

https://doi.org/10.1029/WR005i005p00967

Ollech, D. (2019). seastests: Seasonality Tests.  R package version 0.14.2. https://CRAN.R-project.org/package=seastests

Qiu, D. (2015). aTSA: Alternative Time Series Analysis. R package version 3.1.2. https://CRAN.R-project.org/package=aTSA

R Core Team (2019). R: A language and environment for statistical computing. R Foundation for Statistical Computing, Vienna, Austria. URL https://www.R-project.org/ Version 3.6.1 (2019-07-05)

Saha, K.; Madhavan, V.; Chandrashekhar, G.R. (2020) Pitfalls in long memory research. Cogent Economics and Finance 8: 1733280 https://doi.org/10.1080/23322039.2020.1733280

Shang, H. (2020). A Comparison of Hurst Exponent Estimators in Long-range Dependent Curve Time Series. *Journal of Time Series Econometrics*, *12*(1). https://doi.org/10.1515/jtse-2019-0009

Slino, M.; Scudero, S.; D'Alessandro, A. (2020) Stochastic models for radon daily time series: seasonality, stationarity, and long-range memory detection. Frontiers in Earth Sciences DOI: 10.3389/feart.2020.575001 Viewed on 10/10/2020 at http://hdl.handle.net/2122/13726

Trapletti, A.; Hornik, K. (2019). tseries: Time Series Analysis and Computational Finance. R package version 0.10-47.

Wu, K.; Chen, S.(2020) Long memory and efficiency of Bitcoin under heavy tails, Applied Economics, 52:48, 5298-5309, DOI: 10.1080/00036846.2020.1761942

Appendix: R code for transparency and reproducibility.

# Start: R code with dataset #

#Trademarks
#Go to TESS: http://tmsearch.uspto.gov/bin/gate.exe?f=tess&state=4804:57thz4.1.1
# Manually search and collect Number of Trademarks as follows:
#By Filing Date: "(198712$)[FD]" - Where 198712$ is the %Y%m$.

#Patents
#Go to PATFT: http://patft.uspto.gov/netahtml/PTO/search-adv.htm
#By Application Filing Date: "APD/12/$/2018"

#The patent and trademark filings data were collected from Sept 1977 to Dec 2018.

#Confirm version of R:

> citation()

R Core Team (2019). R: A language and environment for statistical computing. R Foundation for Statistical Computing, Vienna, Austria. URL https://www.R-project.org/.

> version

```
_
platform       x86_64-w64-mingw32
arch           x86_64
os             mingw32
system         x86_64, mingw32
status
major          3
minor          6.1
year           2019
month          07
day            05
svn rev        76782
language       R
version.string R version 3.6.1 (2019-07-05)
nickname       Action of the Toes
```

#Read into R:

> IP<- read.csv("…/data.csv", sep=",")

#Confirm dataframe - length/variables and shrink by 24m to avoid so-called 'patent cliff':

```
> str(IP)
```

'data.frame':   496 obs. of  3 variables:
$ Date: Factor w/ 496 levels "1/1/1978","1/1/1979",..: 455 42 84 126 1 168 209 250 291 332 ...
$ Number.of.Trademark.Applications: int  2669 2597 2552 2604 2386 2370 3126 2738 3028 3088 ...
$ Number.of.Patent.Applications   : int  5760 5898 5731 6630 5064 5439 6660 5799 6487 6419 ...

```
# Shrinking by 24 months dues to so-called "patent-cliff"
> TrademarksTotal<-IP$Number.of.Trademark.Applications[1:472]
> PatentsTotal<-IP$Number.of.Patent.Applications[1:472]

 #Convert to Time-Series, decompose time-series, and perform descriptive statistics
> tsTrademarks<-ts(TrademarksTotal,start=c(1977,9),frequency=12)
> tsPatents<-ts(PatentsTotal,start=c(1977,9),frequency=12)

> tsTrademarks
```

|      | Jan | Feb | Mar | Apr | May | Jun | Jul | Aug | Sep | Oct | Nov | Dec |
|------|-----|-----|-----|-----|-----|-----|-----|-----|-----|-----|-----|-----|
| 1977 |     |     |     |     |     |     |     |     | 2669 | 2597 | 2552 | 2604 |
| 1978 | 2386 | 2370 | 3126 | 2738 | 3028 | 3088 | 2708 | 2638 | 2465 | 2793 | 2636 | 2362 |
| 1979 | 2518 | 2350 | 2920 | 2968 | 2953 | 2794 | 2741 | 2829 | 2438 | 2956 | 2676 | 2596 |
| 1980 | 2469 | 2607 | 3035 | 2893 | 2797 | 3094 | 2883 | 2590 | 2928 | 4081 | 3412 | 3559 |
| 1981 | 3329 | 4113 | 3906 | 4297 | 3871 | 4358 | 4077 | 3815 | 3688 | 3879 | 4162 | 4140 |
| 1982 | 3594 | 4009 | 5128 | 4868 | 4576 | 5244 | 4942 | 5264 | 15843 | 1895 | 3126 | 3529 |
| 1983 | 3915 | 3597 | 4224 | 4297 | 4389 | 4543 | 4192 | 4893 | 4265 | 4634 | 4189 | 4296 |
| 1984 | 4281 | 4472 | 5167 | 5068 | 4762 | 4837 | 4914 | 4803 | 4064 | 4896 | 4922 | 4474 |
| 1985 | 4296 | 4408 | 5087 | 5417 | 5537 | 4914 | 5537 | 5215 | 4627 | 5218 | 4747 | 4904 |
| 1986 | 4785 | 4677 | 5569 | 5153 | 5397 | 5630 | 5152 | 5124 | 5715 | 4787 | 4950 | 4931 |
| 1987 | 4188 | 4826 | 5528 | 5780 | 5372 | 5860 | 6269 | 5990 | 5309 | 5624 | 5454 | 5286 |
| 1988 | 4758 | 5756 | 5730 | 5618 | 6358 | 5706 | 5473 | 6341 | 5776 | 5899 | 5384 | 5744 |
| 1989 | 5546 | 5761 | 6615 | 6230 | 7071 | 6496 | 5995 | 6567 | 5613 | 6687 | 11400 | 8450 |
| 1990 | 9209 | 8797 | 10687 | 9936 | 9836 | 9707 | 9319 | 9634 | 8167 | 9567 | 8417 | 7546 |
| 1991 | 7906 | 7986 | 9192 | 9734 | 9441 | 8962 | 9371 | 9260 | 8884 | 9073 | 8805 | 7972 |
| 1992 | 7486 | 8612 | 10248 | 9940 | 8776 | 10078 | 10019 | 9701 | 9348 | 7853 | 8599 | 9107 |
| 1993 | 9143 | 9143 | 11059 | 11352 | 10594 | 11897 | 11160 | 11628 | 11448 | 10374 | 11248 | 10923 |
| 1994 | 9599 | 10388 | 12173 | 11664 | 12292 | 12902 | 10728 | 12536 | 11443 | 11932 | 11339 | 10634 |
| 1995 | 10961 | 11857 | 14860 | 12391 | 14323 | 13644 | 12325 | 14434 | 12631 | 13726 | 12926 | 12072 |
| 1996 | 12582 | 13872 | 15117 | 15636 | 15565 | 14433 | 15447 | 15590 | 15234 | 16299 | 14727 | 13810 |
| 1997 | 14060 | 15229 | 16962 | 17079 | 16460 | 16465 | 16425 | 15363 | 16089 | 16901 | 14367 | 15163 |
| 1998 | 13510 | 15465 | 17624 | 17224 | 15805 | 17862 | 17515 | 16270 | 16331 | 17426 | 16737 | 16897 |
| 1999 | 16002 | 18431 | 22143 | 20957 | 20515 | 24106 | 21999 | 23891 | 22563 | 24753 | 24225 | 24272 |
| 2000 | 23412 | 25815 | 30423 | 25204 | 26986 | 24831 | 21769 | 24626 | 21644 | 22962 | 20340 | 17822 |
| 2001 | 19248 | 18923 | 20732 | 19781 | 20778 | 18551 | 18311 | 19608 | 14770 | 17727 | 15587 | 14181 |

```
2002 16273 16357 18555 18946 19622 17734 18359 18958 17639 19111 16487 16278
2003 16170 17077 19352 19461 19563 19045 19687 18614 19518 20752 17857 18083
2004 17848 20221 24380 22857 20851 21971 20794 21686 20329 20545 20163 19727
2005 19852 21156 25021 23106 22583 23621 21283 24405 21769 21419 21833 20716
2006 21532 22810 27578 22750 24386 24861 21580 25319 22311 24077 22658 20909
2007 23774 24365 28236 25640 27467 26565 25520 27860 23629 27407 24521 21614
2008 25011 25689 27048 27577 26586 24960 26086 23857 24425 24124 19074 19873
2009 19111 20849 23869 23573 22337 24084 23578 22625 22336 23795 21248 21987
2010 21221 23097 28105 25119 23953 25080 22796 23413 23092 23739 23151 21794
2011 22686 24232 30145 26896 26187 27407 25024 27708 25636 24807 24550 23480
2012 24295 27305 29660 27660 28521 26754 26636 27794 24270 27369 24447 22310
2013 25382 26630 28633 29002 29609 26566 27697 28612 26811 28991 25622 24625
2014 26543 27243 30853 30752 29450 28680 30393 28186 30059 32019 26212 26558
2015 27182 29681 35025 33786 31123 34043 33316 31327 32138 32584 29909 30094
2016 29461 32456 37317 35506 34861 35211 32027 35356 33293 31394 31452 33063
```

> tsPatents

```
      Jan  Feb  Mar  Apr  May  Jun  Jul  Aug  Sep  Oct  Nov  Dec

1977                      5760 5898 5731 6630
1978 5064 5439 6660 5799 6487 6419 5671 5831 5697 6012 5756 6210
1979 5303 5240 6131 6071 6247 6087 5726 5999 5541 6459 5913 6399
1980 5492 5619 6491 5971 6292 6325 6077 5328 6143 6246 5414 6726
1981 4899 5351 6548 5857 5583 6377 5907 5539 5489 5726 5602 6354
1982 4677 5313 6793 5724 5702 6543 5894 5834 10870 3134 4604 5692
1983 4698 4900 6279 5451 5657 6202 5390 5621 5740 5648 5544 6507
1984 4940 5557 6360 6216 6391 6522 6033 6230 5708 6631 6319 6774
1985 5381 6111 6664 6807 6895 6257 6713 6290 6814 7204 6247 7251
1986 5730 6296 7137 7026 6837 6982 6994 6486 7150 7544 6200 7814
1987 5918 6753 7899 7633 7027 7852 7784 6968 7510 7814 7539 8614
1988 6315 7682 8992 7850 8040 8737 7843 8262 8551 8721 8082 9321
1989 7616 8117 9466 8379 9123 9378 8133 8702 8588 9083 8706 9308
1990 8151 8472 9707 9061 9197 9331 8963 9269 8533 10444 8128 9421
1991 7879 8270 9363 9413 9406 9080 9394 8954 9327 9737 9155 9704
1992 8237 8485 10025 9591 8881 10282 9822 8666 10009 9586 9180 10693
1993 8063 8728 11064 9963 9246 10500 9793 9531 10232 9904 10034 11401
1994 9243 9614 12007 10453 10893 12276 10303 11300 12499 11444 11444 13760
1995 10445 10716 13835 11945 17222 28123 8860 10340 11124 10913 11152 12746
1996 10346 11112 13067 12689 13000 13285 13677 13195 14326 14524 13319 15858
1997 13121 13531 15641 15287 15015 16286 15456 14815 16658 16871 14529 16885
1998 12722 13746 16543 15062 14511 16769 15660 14426 15607 15603 14861 18092
1999 12865 14204 17997 15791 15337 17893 15963 16308 17160 16447 16702 19417
```

```
2000 14586 16451 20021 15922 17884 19641 15521 18278 19273 17834 18532 19122
2001 16493 17127 21285 17692 18593 20122 18202 19753 17856 19239 18210 19660
2002 17694 17281 20278 18963 19891 19568 19240 18754 19562 20034 17799 21302
2003 16740 16898 19990 18907 18242 19748 18830 17542 19825 19708 16896 22063
2004 15528 16931 21274 18212 16712 20593 17779 17943 19884 17521 18130 22102
2005 14968 17021 22490 18546 17987 21090 16914 18462 19787 17592 18133 22188
2006 14550 16701 22216 17208 19402 21150 17560 19571 19828 19312 18999 22766
2007 16930 17539 22271 18427 19480 20506 18797 20267 19143 25045 18195 21820
2008 16717 18825 21168 19762 19318 21035 19896 18546 21495 21413 18120 23736
2009 15375 17696 21848 18816 17438 20736 19164 17736 20407 19603 18327 23801
2010 14972 17696 19786 20199 19019 22414 19760 19867 21604 20657 20612 25143
2011 17261 18992 25367 20428 20964 23822 20058 22268 26472 20658 21926 26522
2012 17222 19167 21736 22546 23921 25237 23054 24756 27051 22929 24437 27708
2013 18316 23728 42788 19501 22634 22932 23350 23313 25042 25126 23015 28838
2014 19391 22253 30969 23387 24040 25581 24700 23086 27091 25141 21855 29294
2015 19599 21643 24767 23184 22546 26852 23783 22590 25856 23159 21726 27049
2016 18970 20933 23298 19948 20837 22912 18417 20752 21299 17774 17791 19436
```

```
> plot(decompose(tsTrademarks,type="additive"))
> plot(decompose(tsPatents,type="additive"))
```

```
#Identify outliers
#Javier López-de-Lacalle (2019). tsoutliers: Detection of Outliers in Time Series. R package version 0.6-8.
# https://CRAN.R-project.org/package=tsoutliers
library(tsoutliers)
> TrademarksOutliers<-tso(tsTrademarks,types = c("AO","LS","TC"),maxit.iloop=10)
> PatentsOutliers<-tso(tsPatents,types = c("AO","LS","TC"),maxit.iloop=10)
```

```
> TrademarksOutliers
Series: tsTrademarks
Regression with ARIMA(2,1,1)(0,1,2)[12] errors
Coefficients:
         ar1     ar2     ma1     sma1     sma2      AO61      LS147     LS262
     -1.0107  -0.5826  0.4306  -0.4939  -0.2779  12137.5320  4527.2969  3409.3950
s.e.  0.0834   0.0440  0.1067   0.0480   0.0458    669.2913   681.1637   674.9746
sigma^2 estimated as 868751:  log likelihood=-3790.79
AIC=7599.58   AICc=7599.98   BIC=7636.74
```

```
Outliers:
  type ind   time    coefhat  tstat
1  AO   61 1982:09   12138   18.135
2  LS  147 1989:11    4527    6.646
3  LS  262 1999:06    3409    5.051
```

```
> PatentsOutliers
Series: tsPatents
Regression with ARIMA(3,0,0)(2,1,2)[12] errors
Coefficients:
       ar1     ar2     ar3    sar1     sar2     sma1    sma2      AO61       AO214
    0.2731  0.2776  0.4185  0.7318  -0.3230  -1.4584  0.6303  5591.8986  15515.5416
s.e. 0.0480  0.0438  0.0468  0.1133   0.0673   0.1146  0.0879   773.2661    764.2898
        AO362       AO427
     5058.3040  17555.5353
s.e.  757.5541    799.8064
sigma^2 estimated as 917822:  log likelihood=-3812.79
AIC=7649.58   AICc=7650.27   BIC=7699.15

Outliers:
  type ind   time coefhat  tstat
1  AO  61 1982:09   5592  7.232
2  AO 214 1995:06  15516 20.301
3  AO 362 2007:10   5058  6.677
4  AO 427 2013:03  17556 21.950

> plot(TrademarksOutliers); X11(); plot(PatentsOutliers)

#Clean/smooth data - replace identified outliers (X) with average of prior (X(t-1)) and posterior (X(t+1))

>Trademarks<-tsTrademarks; Patents<-tsPatents
>Trademarks[61]= (Trademarks[62]+Trademarks[64]) / 2
>Trademarks[147]= (Trademarks[146]+Trademarks[148]) / 2
>Trademarks[262]= (Trademarks[261]+Trademarks[263]) / 2
>Patents[61]= (Patents[62]+Patents[64]) / 2
>Patents[214]= (Patents[213]+Patents[215]) / 2
>Patents[362]= (Patents[361]+Patents[363]) / 2
>Patents[427]= (Patents[426]+Patents[428]) / 2

> plot(decompose(Trademarks,type="additive"))
> plot(decompose(Patents,type="additive"))

> library(moments); citation("moments")

#Lukasz Komsta and Frederick Novomestky (2015). moments: Moments, cumulants, skewness, kurtosis
#and related tests. R package version 0.14. https://CRAN.R-project.org/package=moments

#Use fitted output from tsoutliers

>summary(Trademarks); sd(Trademarks); skewness(Trademarks); kurtosis(Trademarks)
Min. 1st Qu.  Median    Mean 3rd Qu.    Max.
```

1895   5537   15456   15276   23574   37317
[1] 9418.054
[1] 0.2020644
[1] 1.763723
>summary(Patents); sd(Patents); skewness(Patents); kurtosis(Patents)
Min. 1st Qu.  Median    Mean 3rd Qu.    Max.
3134   7598   14468   13930   19422   30969
[1] 6523.347
[1] 0.1854975
[1] 1.757654

#note: skew/kurtosis comparative - no need to transform

#auto/serial correlation

acf(Trademarks);pacf(Trademarks)

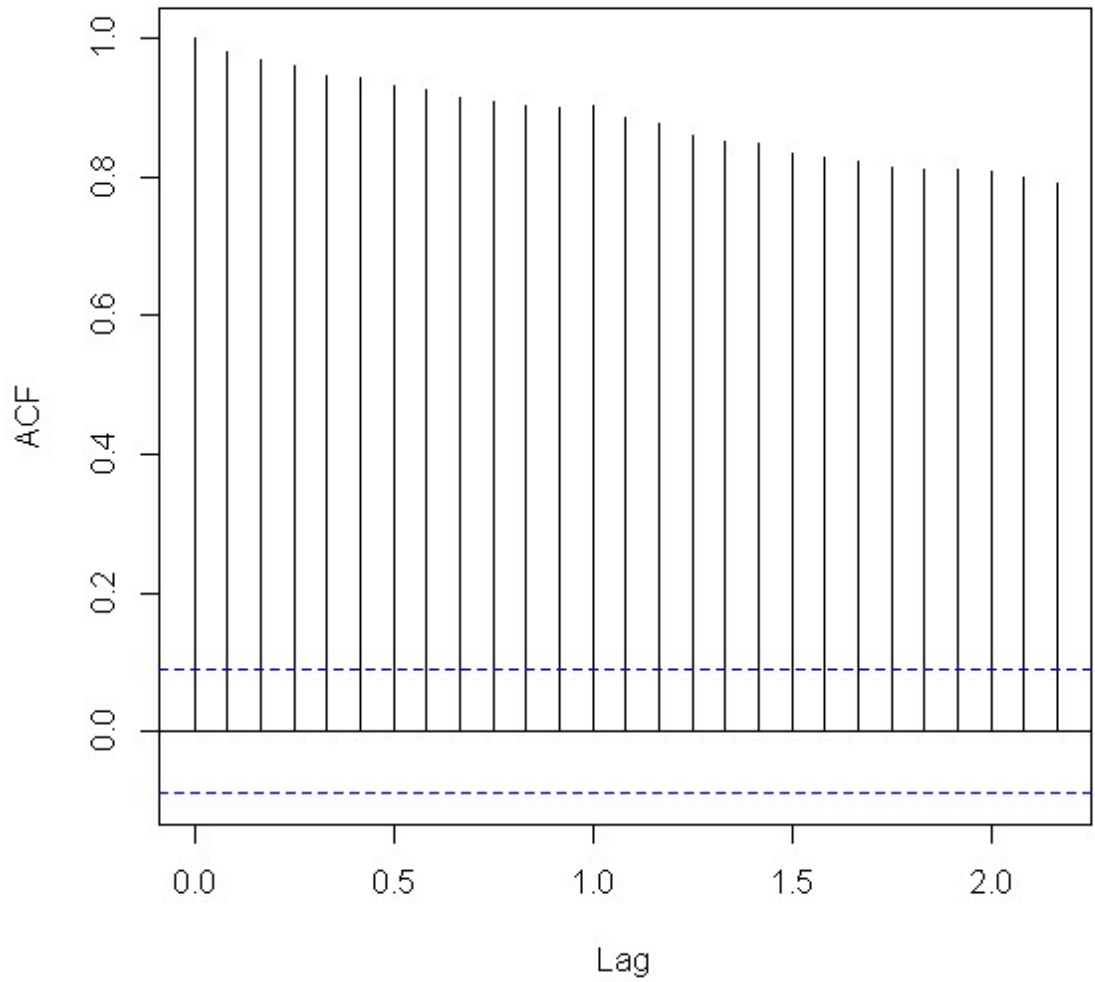

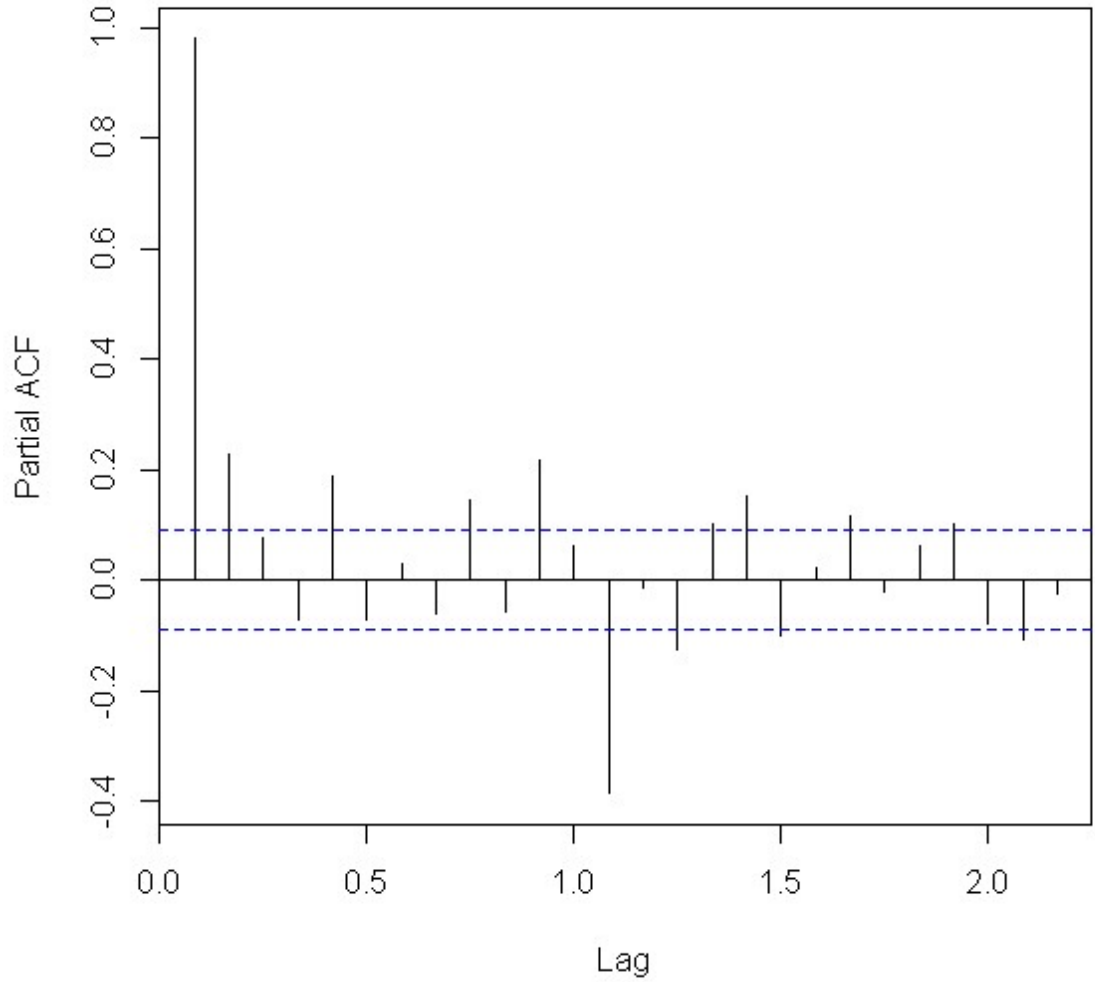

acf(Patents);pacf(Patents)

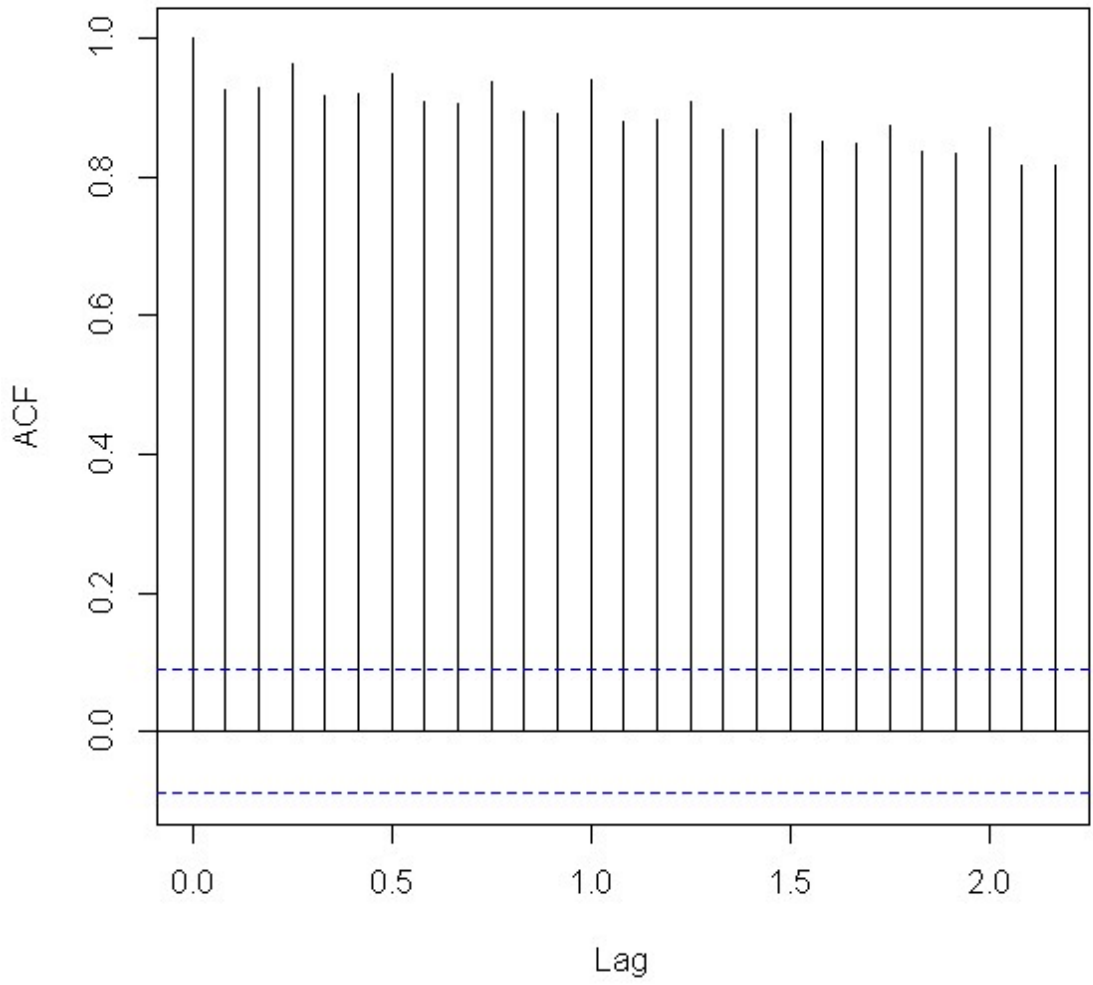

**Series  Patents**

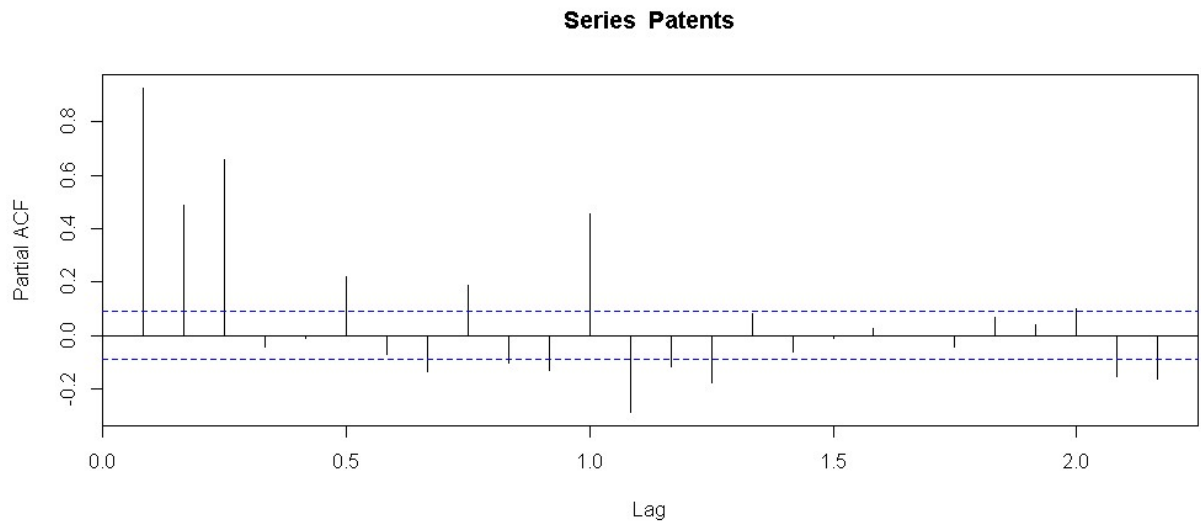

#Perform normality, stationarity, seasonality, long-memory, and non-linearity tests
> #normality test

> library(nortest);citation("nortest")

To cite package 'nortest' in publications use: Juergen Gross and Uwe Ligges (2015). nortest: Tests for Normality. R package version 1.0-4. https://CRAN.R-project.org/package=nortest

> ad.test(Trademarks) #null normality

     Anderson-Darling normality test
data:  Trademarks
A = 11.055, p-value < 2.2e-16

> ad.test(Patents) #null normality

     Anderson-Darling normality test
data:  Patents
A = 13.102, p-value < 2.2e-16

> cvm.test(Trademarks)

     Cramer-von Mises normality test
data:  Trademarks
W = 1.7112, p-value = 7.37e-10

Warning message:
In cvm.test(Trademarks) : p-value is smaller than 7.37e-10, cannot be computed more accurately

> cvm.test(Patents)

Cramer-von Mises normality test
data:  Patents
W = 2.2197, p-value = 7.37e-10

Warning message: In cvm.test(Patents) : p-value is smaller than 7.37e-10, cannot be computed more accurately

```
> #stationary test

>library(forecast)
>ndiffs(Trademarks, test= "kpss"); ndiffs(Trademarks, test= "adf"); ndiffs(Trademarks, test= "pp")
[1]  1
[1]  1
[1]  1
>ndiffs(Patents, test= "kpss"); ndiffs(Patents, test= "adf"); ndiffs(Patents, test= "pp")
[1]  1
[1]  1
[1]  1

> library(aTSA)

Attaching package: 'aTSA'
The following object is masked from 'package:forecast':forecast
The following object is masked from 'package:graphics': identify

> citation("aTSA")
To cite package 'aTSA' in publications use: Debin Qiu (2015). aTSA: Alternative Time Series Analysis. R
package version 3.1.2.  https://CRAN.R-project.org/package=aTSA

> stationary.test(Trademarks, method="kpss")

KPSS Unit Root Test alternative: nonstationary

Type 1: no drift no trend
lag stat p.value
5 0.59    0.1
-----

Type 2: with drift no trend
lag  stat p.value
5 0.751    0.01
-----

Type 1: with drift and trend
lag  stat p.value
```

5 0.181   0.023
-----------
Note: p.value = 0.01 means p.value <= 0.01: p.value = 0.10 means p.value >= 0.10

> stationary.test(Patents, method="kpss")

KPSS Unit Root Test alternative: nonstationary

Type 1: no drift no trend
lag stat p.value
5 5.88   0.01
-----
Type 2: with drift no trend
lag stat p.value
5 5.53   0.01
-----
Type 1: with drift and trend
lag  stat p.value
5 0.524   0.01
-----------
Note: p.value = 0.01 means p.value <= 0.01: p.value = 0.10 means p.value >= 0.10

#Now the first diffs

stationary.test(diff(Trademarks), method="kpss")

KPSS Unit Root Test alternative: nonstationary

 Type 1: no drift no trend
lag  stat p.value
5 0.642   0.1
-----
 Type 2: with drift no trend
lag   stat p.value
5 0.0294   0.1
-----
Type 1: with drift and trend
lag  stat p.value
5 0.015   0.1
-----------

Note: p.value = 0.01 means p.value <= 0.01: p.value = 0.10 means p.value >= 0.10

stationary.test(diff(Patents), method="kpss")
KPSS Unit Root Test

alternative: nonstationary
Type 1: no drift no trend
lag  stat p.value
5 0.799    0.1
-----
Type 2: with drift no trend
lag   stat p.value
5 0.0858    0.1
-----
Type 1: with drift and trend
lag   stat p.value
5 0.0716    0.1
-----------
Note: p.value = 0.01 means p.value <= 0.01 : p.value = 0.10 means p.value >= 0.10

#Long memory

> library(LongMemoryTS); citation("LongMemoryTS")
To cite package 'LongMemoryTS' in publications use: Christian Leschinski (2019). LongMemoryTS: Long
Memory Time Series. R package version  0.1.0. https://CRAN.R-project.org/package=LongMemoryTS

#Qu Test
> Qu.test(diff(Trademarks),m)
$W.stat [1] 1.432811

$CriticalValues

|  | eps=.02 | eps=.05 |
|---|---|---|
| alpha=.1 | 1.118 | 1.022 |
| alpha=.05 | 1.252 | 1.155 |
| alpha=.025 | 1.374 | 1.277 |
| alpha=.01 | 1.517 | 1.426 |

> Qu.test(diff(Patents),m)
$W.stat [1] 2.334971

$CriticalValues

|  | eps=.02 | eps=.05 |
|---|---|---|
| alpha=.1 | 1.118 | 1.022 |
| alpha=.05 | 1.252 | 1.155 |
| alpha=.025 | 1.374 | 1.277 |
| alpha=.01 | 1.517 | 1.426 |

```
> #multivariate local Whittle Score

> MLWS(diff(Trademarks), m=m)

$B
     [,1]
[1,]   1

$d [1] -0.3782234

$W.stat [1] 1.502709

$CriticalValues

 alpha=.1  alpha=.05 alpha=.025  alpha=.01
1.118     1.252     1.374     1.517

> MLWS(diff(Patents), m=m)

$B
     [,1]
[1,]   1

$d [1] -0.3440901

$W.stat [1] 2.43958

$CriticalValues

 alpha=.1  alpha=.05 alpha=.025  alpha=.01
1.118     1.252     1.374     1.517

> #seasonality tests

> library(seastests); citation("seastests")
```

To cite package 'seastests' in publications use: Daniel Ollech (2019). seastests: Seasonality Tests. R package version 0.14.2. https://CRAN.R-project.org/package=seastests

```
> summary(wo(Trademarks))

Test used:  WO
Test statistic:  1
P-value:  0 0 0

The WO - test identifies seasonality
```

```
> summary(wo(Patents))
```

Test used:  WO
Test statistic:  1
P-value:  0 0 0

#

The WO - test identifies seasonality

```
> isSeasonal(Trademarks,"qs"); isSeasonal(Trademarks,"fried"); isSeasonal(Trademarks,"welch");
```

[1] TRUE
[1] TRUE
[1] TRUE

```
> isSeasonal(Patents,"qs"); isSeasonal(Patents,"fried"); isSeasonal(Patents,"welch");
```

[1] TRUE
[1] TRUE
[1] TRUE

#Nonlinearity Tests

```
> library(nonlinearTseries); citation("nonlinearTseries")
```

Attaching package: 'nonlinearTseries' The following object is masked from 'package:aTSA': estimate

The following object is masked from 'package:grDevices':  contourLines

To cite package 'nonlinearTseries' in publications use: Constantino A. Garcia (2020). nonlinearTseries: Nonlinear Time Series Analysis. R package version 0.2.10. https://CRAN.R-project.org/package=nonlinearTseries

```
> nonlinearityTest(Trademarks)
```

     ** Teraesvirta's neural network test  **
     Null hypothesis: Linearity in "mean"
     X-squared = 13.65589  df = 2  p-value = 0.001083081

     ** White neural network test  **
     Null hypothesis: Linearity in "mean"
     X-squared = 14.43927  df = 2  p-value = 0.00073207

     ** Keenan's one-degree test for nonlinearity  **
     Null hypothesis: The time series follows some AR process
     F-stat =  0.04386686  p-value =  0.8342031

** McLeod-Li test  **
Null hypothesis: The time series follows some ARIMA process
Maximum p-value =  0

** Tsay's Test for nonlinearity **
Null hypothesis: The time series follows some AR process
F-stat =  3.041166  p-value =  8.100652e-10

** Likelihood ratio test for threshold nonlinearity **
Null hypothesis: The time series follows some AR process
Alternativce hypothesis: The time series follows some TAR process
X-squared =  51.55873  p-value =  0.03278644

> nonlinearityTest(Patents)

** Teraesvirta's neural network test  **
Null hypothesis: Linearity in "mean"
X-squared =  110.8611  df =  2  p-value =  0

** White neural network test  **
Null hypothesis: Linearity in "mean"
X-squared =  108.337  df =  2  p-value =  0

** Keenan's one-degree test for nonlinearity  **
Null hypothesis: The time series follows some AR process
F-stat =  3.379455  p-value =  0.06672325

** McLeod-Li test  **
Null hypothesis: The time series follows some ARIMA process
Maximum p-value =  0

** Tsay's Test for nonlinearity **
Null hypothesis: The time series follows some AR process
F-stat =  6.296795  p-value =  1.162674e-15

** Likelihood ratio test for threshold nonlinearity **
Null hypothesis: The time series follows some AR process
Alternative hypothesis: The time series follows some TAR process
X-squared =  78.45917  p-value =  2.60603e-05

```
> library(tsfeatures); citation("tsfeatures")
```

To cite package 'tsfeatures' in publications use: Rob Hyndman, Yanfei Kang, Pablo Montero-Manso, Thiyanga Talagala, Earo Wang, Yangzhuoran Yang and Mitchell O'Hara-Wild (2020). tsfeatures: Time Series Feature  Extraction. R package version 1.0.2. https://CRAN.R-project.org/package=tsfeatures

```
hurstTrademarks=0;hurstPatents=0
endT<-length(Trademarks); endP<-length(Patents)
for (i in 1:endT) { hurstTrademarks[i] <- hurst (Trademarks[1:(1+i*1)]) }
for (i in 1:endP) { hurstPatents[i] <- hurst (Patents[1:(1+i*1)]) }

hurstTrademarks<-ts(hurstTrademarks,start=c(1977,9),end=c(2016,12), frequency=12)
hurstPatents<-ts(hurstPatents,start=c(1977,9),end=c(2016,12), frequency=12)
plot(hurstTrademarks); plot(hurstPatents)
```

```
> library(tseries);citation("tseries")
```

'tseries' version: 0.10-47 'tseries' is a package for time series analysis and computational finance. See 'library(help="tseries")' for details.

Attaching package: 'tseries' The following objects are masked from 'package:aTSA': adf.test, kpss.test, pp.test

To cite in publications use:

Adrian Trapletti and Kurt Hornik (2019). tseries: Time Series Analysis and Computational Finance. R package version 0.10-47.

```
plot(ts.intersect(Patents, hurstPatents),main="", yax.flip=TRUE)
```

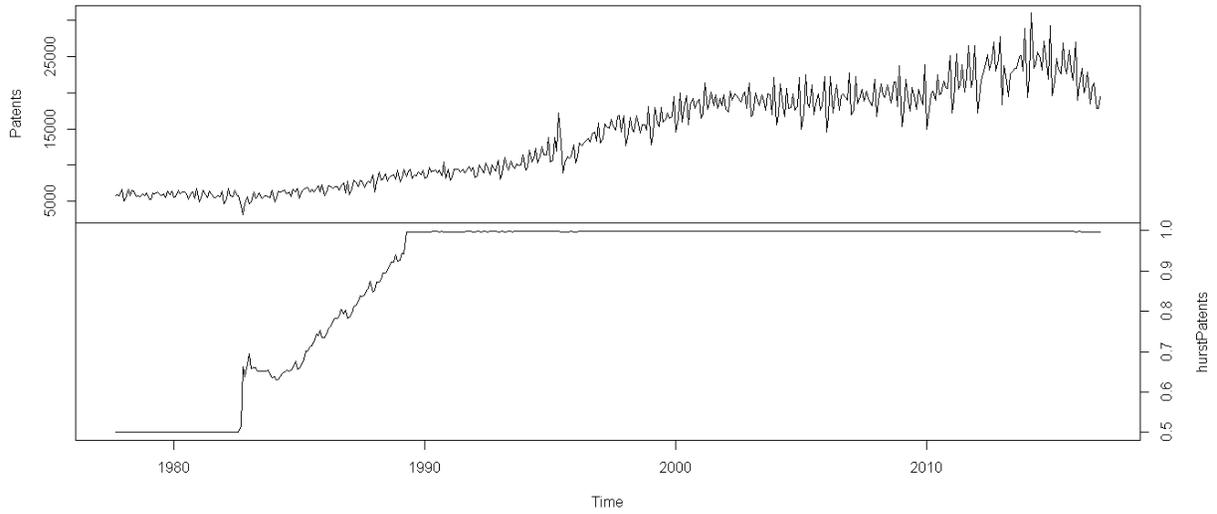

plot(ts.intersect(Trademarks, hurstTrademarks),main="", yax.flip=TRUE)

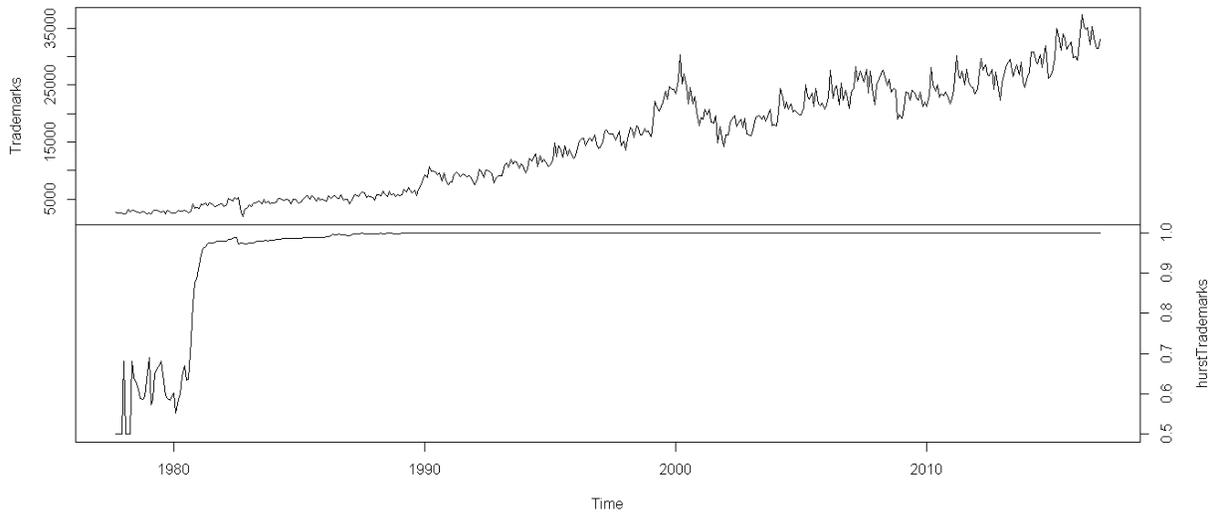

plot(ts.intersect(hurstPatents, hurstTrademarks),main="", yax.flip=TRUE)

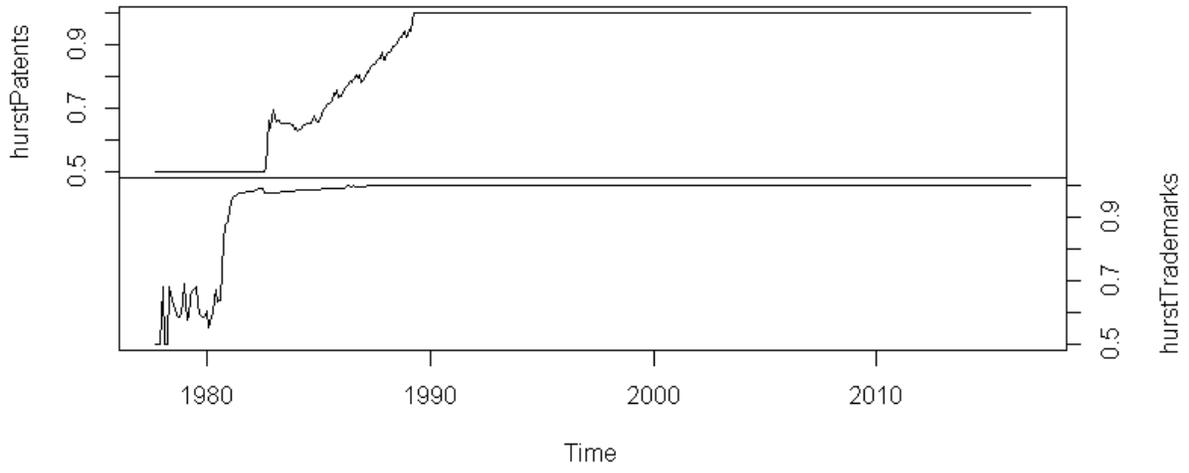

#### End ####